# Transforming 2D carbon allotropes into 3D ones through topological mapping: The case of biphenylene carbon (graphenylene)

Raphael M. Tromer[1,2], Levi C. Felix[1,2], Ray H. Baughmann[3], Douglas S. Galvao[1,2], and Cristiano F. Woellner[4]

[1]Applied Physics Department, State University of Campinas, Campinas/SP, 13083-970, Brazil

[2]Center for Computational Engineering & Sciences - CCES, State University of Campinas, Campinas/SP, 13083-970, Brazil.

[3]Alan G. MacDiarmid NanoTech Institute, University of Texas at Dallas, Richardson, TX, 75080 USA

[4]Physics Department, Federal University of Paraná - UFPR, Curitiba/PR, 81531-980, Brazil.

**Abstract**

In this work, we propose a new methodology for obtaining 3D carbon allotrope structures from 2D ones through topological mapping. The idea is to select a 3D target structure and 'slice' it along different structural directions, creating a series of 2D structures. As a proof of concept, we chose the Tubulane structure 12-hexa(3,3) as a target. Tubulanes are 3D carbon allotropes based on cross-linked carbon nanotubes. One of obtained 2D 'sliced' structures was mapped into the biphenylene carbon (BPC). We showed that compressing BPC along different directions can generate not only the target Tubulane 12-hexa(3,3) but at least two other structures, bcc-C6 and an unreported member of the Tubulane family, which we called Tubulane X. The methodology proposed here is completely general, it can be used coupled with any quantum method. Considering that new 2D carbon allotropes, such as the biphenylene carbon network, which is closely related to BPC, have been recently synthesized, the approach proposed here opens new perspectives to obtain new 3D carbon allotropes from 2D structures.

# 1. Introduction

The experimental realization of single-layer graphene [1] created a revolution in materials science. Graphene is a 2D carbon allotrope with unique electronic and mechanical properties that have been exploited in many applications [2–6].

The advent of graphene renewed the interest in other 2D carbon allotropes materials and structures that were proposed before graphene, such as biphenylene networks (BPN) [7,8] and graphynes [9], were recently synthesized [10,11]. It also stimulated the search for other non-carbon 2d materials [12]. However, similarly to graphene, most of these 2D materials are obtained from lamellar-like structures (the so-called van der Waals solids) [13]. Recently, the first 2D material from non-van der Waals solids, named hematene, was obtained from liquid exfoliation [14] of 3D hematite. Following the same approach, several new 2D structures from 3D ones have been reported [15].

A natural question is about the inverse process, i.e., how to obtain 3D structures from 2D ones. Of particular interest would be to obtain new 3D structures from 2D ones already experimentally realized. In fact, this is not a new idea, we have the case of diamond (a 3D structure) being obtained from graphite layers (2D structures) under high pressure and/or high temperature [16–19]. However, topologically mapping 2D structures into 3D ones is not a trivial problem, and there are only a few examples reported in the literature [20,21].

In this work, we proposed a new theoretical approach to obtaining 3D carbon allotropes from 2D ones. The idea is to select the 3D target structure and slice it along different structural directions, creating a series of 2D structures. These 2D structures are then fully geometrically optimized and topologically mapped into existing or theoretically proposed 2D carbon allotropes.

As a proof of concept, we chose the Tubulane structure 12-hexa(3,3) as a target, see **Figure 1**. Tubulanes are 3D carbon allotropes based on cross-linked carbon nanotubes [22]. Their synthesis remains elusive up to now.

In **Figure 2** we present (top (**Figure 2(a)**) and lateral views (**Figure 2(b)**)) a supercell of the 12-hexa(3,3) containing 3 layers (indicated by different colors), top (**Figure 2(a)**). In **Figure 2(c)** we present one of the selected sliced 2D structures. **Figure 2(d)-(f)** shows representative snapshots of the optimization process (see Methodology Section). In **Figure 2(f),** we present the optimized structure (top and lateral views). Interestingly, this structure can be mapped into the biphenylene carbon (BPC), one of the structures of the biphenylene network family [10]. BPC was proposed by Baughman and collaborators in 1987 [9]. We then analyzed the topological transformations that can generate 3D structures from 2D BPC.

## 2. Materials and Methods

In order to obtain the 3D carbon allotrope structures from 2D BPC layers, we consider a mechanical-chemistry-like process, following the three steps shown in **Figure 3** (a more detailed scheme is presented in **Figure S1** of the Supplementary Materials. First, we consider a BPC supercell of three overlapping layers initially separated by 2 Angstroms to prevent any covalent chemical bonds before geometry optimizations.

All geometry optimizations were carried out using the semi-empirical PM6-DH2 Hamiltonian (including van der Waals corrections), as implemented in the MOPAC2016 code

[23,24]. We choose as convergence criterion for geometry optimization when the gradient is smaller than 0.1.

Once the system is geometrically optimized, we apply a biaxial (simultaneously) strain along the x and y-directions until the layers chemically react, forming a 3D structure. The compression value/rate can be arbitrarily chosen, corresponding to steps 2 and 3 in **Figure 3**. Different biaxial strain values (initial conditions) applied to 2D layers can, in principle, lead to different 3D structures. After the compression along the x and y-directions is completed, the compression along the z-direction is also applied (step 3 in **Figure 3**). Before this compression, the system is geometrically relaxed along the z-direction.

As the layers are compressed, they deform and can react, forming intralayer and interlayer covalent bonds (this can occur in steps 2 and 3). Once a well-defined 3D structure is formed (step 4 in **Figure 3**), the compression process is stopped, and the obtained 3D structure is then fully optimized (lattice vectors and atom positions, with no constraints).

If the obtained 3D structure is not a defined target, the process can be repeated using different values of the applied strain values and rate compression.

## 3. Results and Discussions

Following the steps shown in **Figure 3**, we used a biaxial strain of 2% along the xy plane, followed by a compression of 10% along the z-direction (in steps of 0.5 Angstroms), we then observed a formation of a well-defined 3D structure. During the structural transformations, the number of atoms is kept constant; we considered 144 carbon atoms in the supercell for all cases discussed in this work. When the system is compressed along the z-direction, the layers are displaced along the plane XY, breaking the initial coupling AA, as shown in step 3 from **Figure 4**). In **Figure 4**, we present a series of representative snapshots of the process. The analysis of the structure (see Table 1) showed that the obtained 3D structure was not the expected tubulane target but a well-known carbon allotrope known as bcc-C6 [25]. The MOPAC prediction for the formation energy of the resultant 3D structure (bcc-C6) is -8.7 eV/atom, which is exactly the same value from DFT calculations [25].

As the obtained 3D structure was not the defined target, we then repeated the process, considering the same initial conditions but increasing the compression along the xy plane. After the system converges to an applied compression of 2% along the xy plane (see step 2 from **Figure 2**)), it is compressed again by 2%, and the process is repeated until the total compression is 10%, as shown in step 2 of **Figure 5**. The layer curvature now is larger than in the previous case, which will be reflected in different chemical reactivity.

The system is then compressed by 10% along the z-direction in steps of 0.5 Angstroms (step 3 from **Figure 5**)). Again, a well-formed 3D structure is observed. The crystallographic analysis (see Table 1) shows that it is the target tubulane 12-hexa(3,3) [22] (see step 4 from **Figure 5**). Unaware of the tubulane work, this structure was 'rediscovered' years later and called bct [26].

Another parameter that affects the resulting 3D structure is the compression rate along the z-direction. For different rate values, different 3D systems are obtained. In **Figure S2** of the Supplementary Materials, we present the results for the structure obtained using the same procedure to obtain the tubulane 12-hexa(3,3) but changing the z-compression rate to a step of 0.8 Angstroms. We could not identify this 3D structure from the literature, but it is extremely similar

(see **Table 1**) to the structures of the tubulane family, but not one of the structures listed in the original tubulane paper; we named it Tubulane X.

In **Figure 6**, we present the formation energy as a function of the simulation steps for the cases discussed above. Initially, we have the compression along the xy plane, followed by successive compressions along the z-direction (red squares). From this Figure, we can see that for all cases, the obtained 3D structures are more stable than their 'parent' 2D ones. The whole process can be better understood from the **videos** in the Supplementary Material.

Up to now, we discussed the procedures to transform the 2D BPC into (at least) three different 3D structures. Further validation of these topological transformations is to carry out the inverse process, i. e., 'slicing' bcc-C6, Tubulane 12-hexa(3,3), and Tubulane X, to recover the 2D BCC structure. We carried out this process and the obtained sliced 2D structures were then geometrically optimized. As expected, the 2D BPC structure is reobtained (see **Figures S3**, **S4**, and **S5** in the Supplementary Materials), thus further validating our topological approach.

**4. Summary and Conclusions**

In this work, we propose a new methodology for obtaining 3D carbon allotrope structures from 2D ones through topological mapping. The idea is to select a 3D target structure and 'slice' it along different structural directions, creating a series of 2D structures. These 2D structures are then fully geometrically optimized (we used the quantum Hamiltonian PM6-DH2, as available in the MOPAC code [23]) and topologically mapped into existing or theoretically proposed 2D carbon allotropes. As proof of concept, we chose the tubulane structure 12-hexa(3,3) as a target. Tubulanes are 3D carbon allotropes based on cross-linked carbon nanotubes [22]. One of the obtained 2D 'sliced' structures was mapped into the biphenylene carbon (BPC). Initially, the BPC compression process did not yield the target structure but the bcc-C6. Then using different parameters, the target structure (Tubulane 12-hexa(3,3)) was obtained, as well as an unreported member of the Tubulane family, which we called Tubulane X. For completeness, we carried the 'reverse test', 'slicing' again the obtained 3D structures (bcc-C6, Tubulane 12-hexa(3,3), and Tubulane X), and their 'parent' 2D BCC was reobtained for all cases. The methodology proposed here is completely general and it can be used coupled with any quantum method. Considering that new 2D carbon allotropes, such as the biphenylene carbon network, which is closely related to BPC, have been recently synthesized (as well as other related structures, such as graphynes and 2D fullerene networks), the approach proposed here opens new perspectives to obtain new 3D carbon allotropes from 2D structures.

## Acknowledgements

This work was financed in part by the Coordenação de Aperfeiçoamento de Pessoal de Nível Superior - Brasil (CAPES) - Finance Code 001 and CNPq and FAPESP. The authors thank the Center for Computational Engineering and Sciences at Unicamp for financial support through the FAPESP/CEPID Grant #2013/08293-7 and Grant #2018/11352-9.

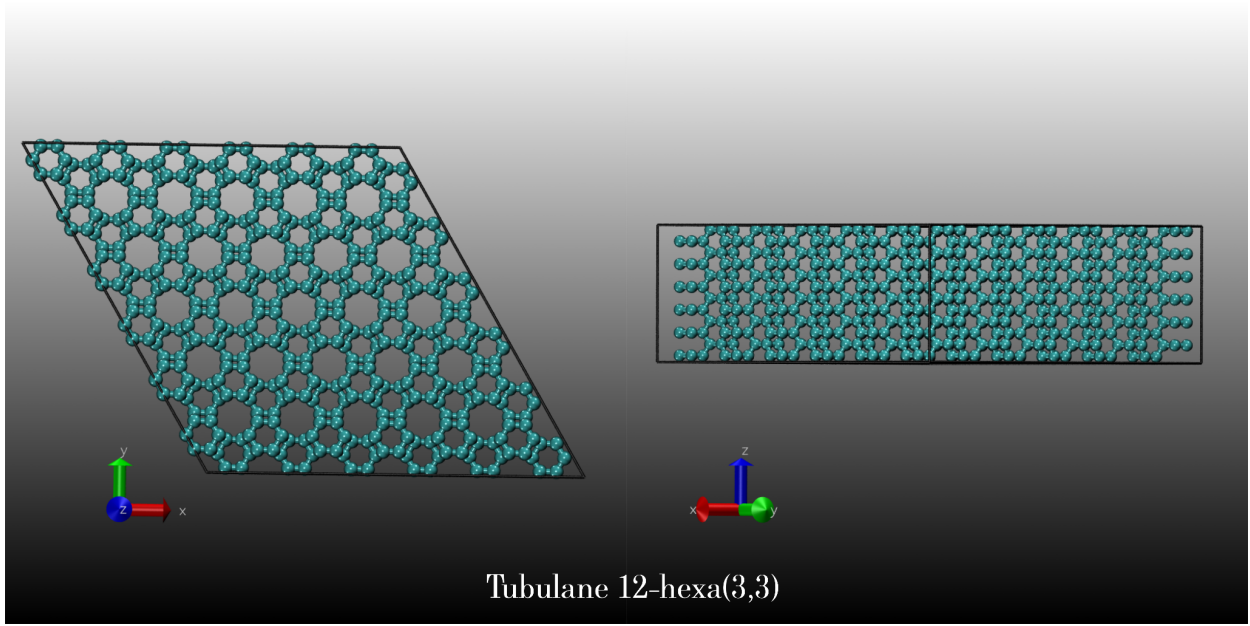

**Figure 1.** Left and right, the top and lateral views of the Tubulane 12-hexa(3,3), the target structure.

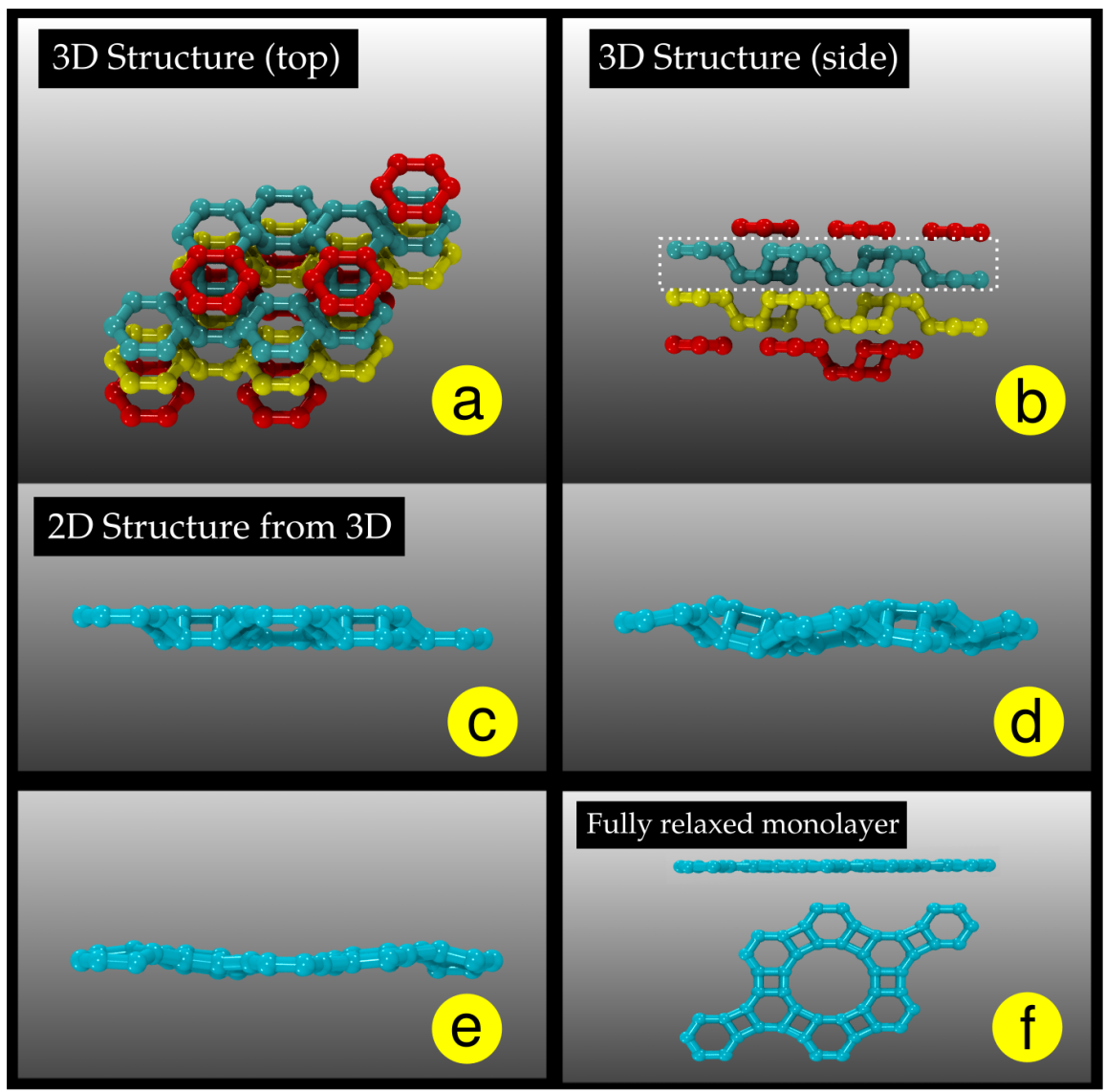

**Figure 2.** (a) Top view of the target structure, Tubulane 12-hexa(3,3). For a better view, the interlayer bond is made transparent; (b) lateral view of (a) selected 'sliced' layers. The dashed rectangle indicates one of the possible 'sliced' 2D structures; (c)-(e) different stages of the geometry optimizations of (b); (f) lateral and top views of the optimized structure, identified as biphenylene carbon (BPC) (graphenylene) [9].

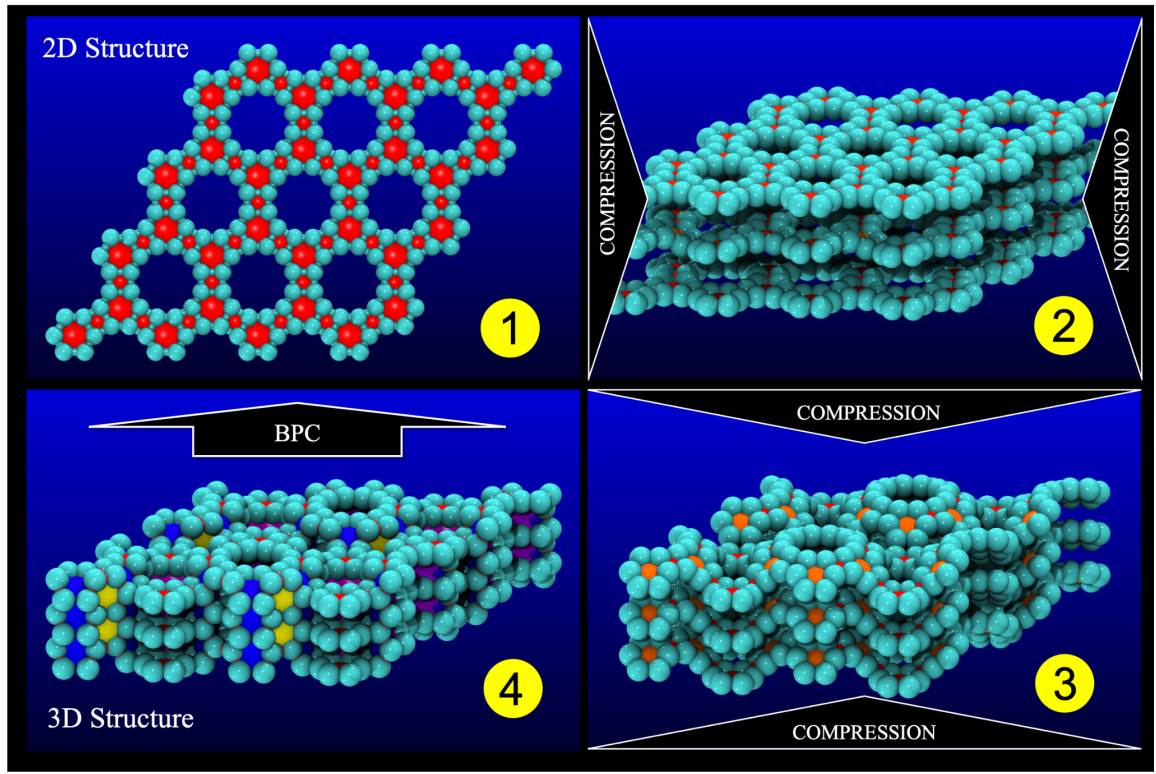

**Figure 3**: Schematic representation of how the 3D structures are obtained from 2D ones. The optimized BPC structure (1) is stacked (2) and compressed along the x, y, and z-directions (2-3) until a stable 3D crystalline structure is obtained (4).

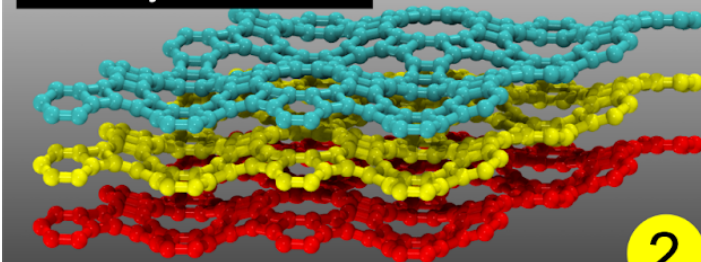
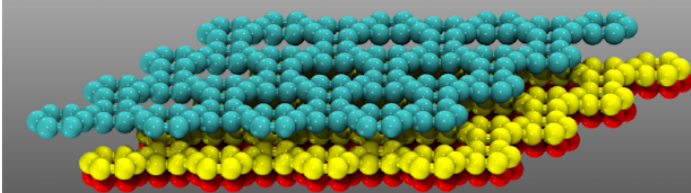
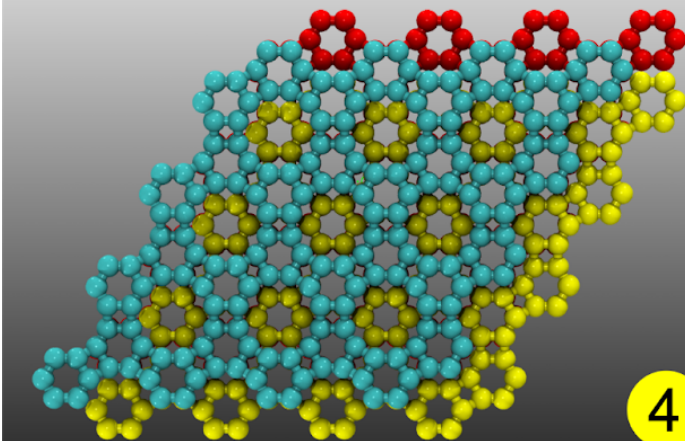
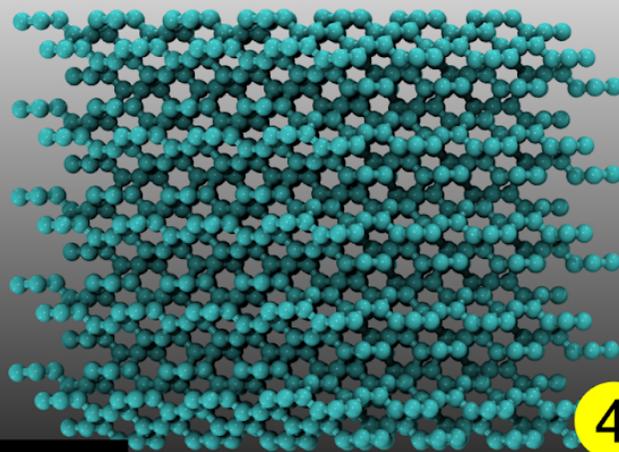

**Figure 4:** Representative snapshots of the structural stages from 2D to 3D structures. Compressed corrugated (buckled) stacked BPC (2). The layers slide and become less bucked (3). Top and lateral views of obtained 3D structure, identified as bcc-C6 [21].

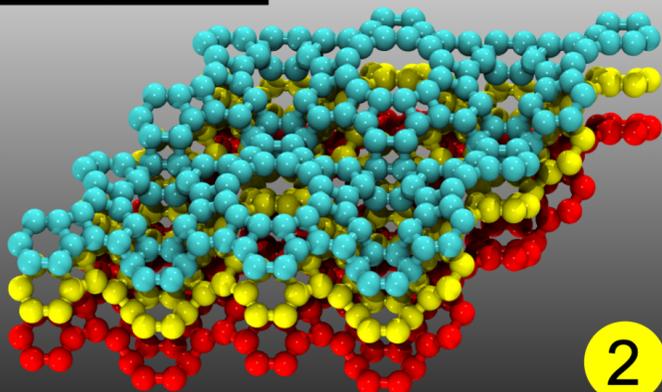

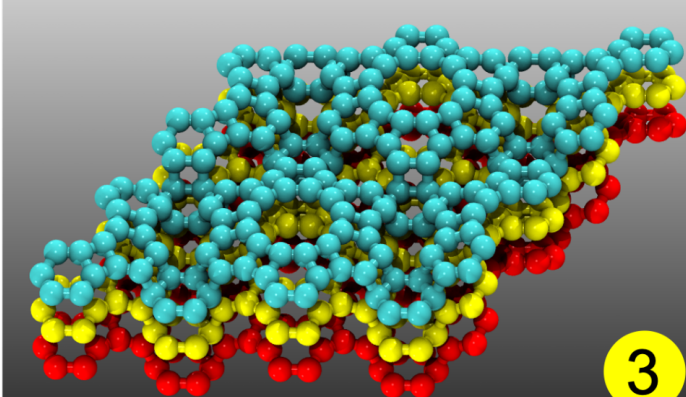

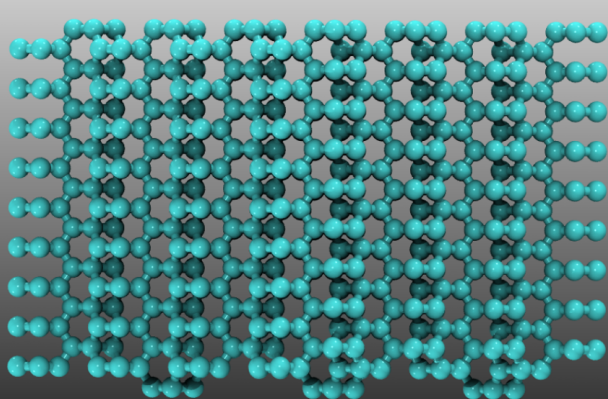

**Figure 5**. Representative snapshots of the structural stages from 2D to 3D ones. Compressed corrugated (buckled) stacked BPC (2). Top (3) and lateral view (4) of obtained 3D structure, identified as the target Tubulane 12-hexa(3,3) [22].

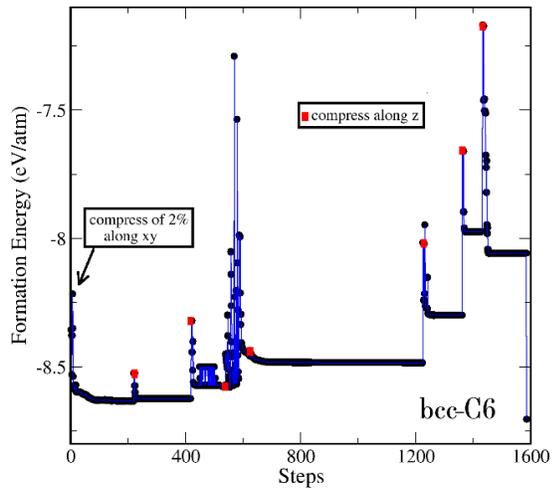

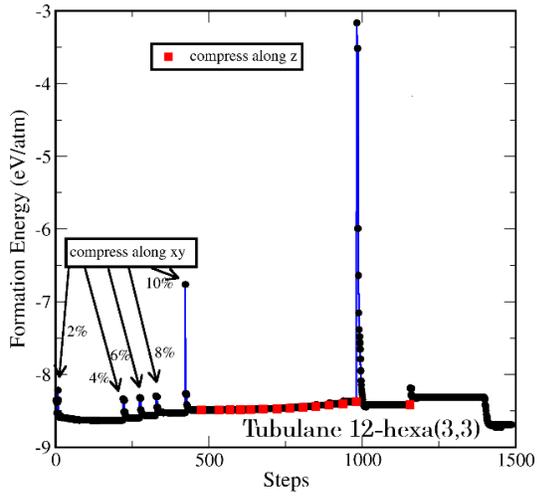

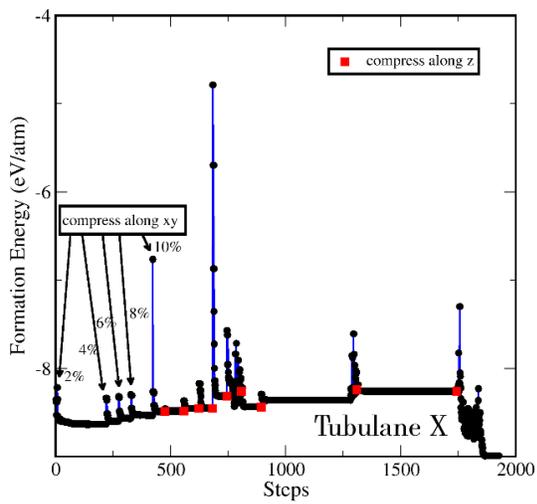

**Figure 6:** Energy profiles during the structural transformation processes from 2D to 3D structures. Top to bottom, bcc-C6, Tubulane 12-hexa(3,3), and Tubulane X. As can be seen from this Figure, although some of the intermediate structural transformations generate 3D structures with higher energy (indicated by the peaks) than the 2D ones, all final optimized 3D structures have lower energy.

| Structures | atoms | Space group | Unit Cell optimized lattice parameters |
|---|---|---|---|
| **BPC [27]** | 48 | P6/mmm (191) | a=b=6.68<br>c=20.00<br>α=β=90<br>ɣ=60 |
| **bcc-C6 [25]** | 144 | P-3m1 (164) | a=b=13.10<br>c=6.70<br>α=β=90<br>ɣ=60 |
| **Tubulane 12-hexa(3,3) [22]** | 144 | P63/mmc (194) | a=b=12.09<br>c=7.79<br>α=β=90<br>ɣ=60 |
| **Tubulane X** | 144 | P6/mmm (191) | a=b=12.70<br>c=7.90<br>α=β=90<br>ɣ=60 |

**Table 1**. Summary of the structural information of the obtained 3D structures and their parent 2D BPC.

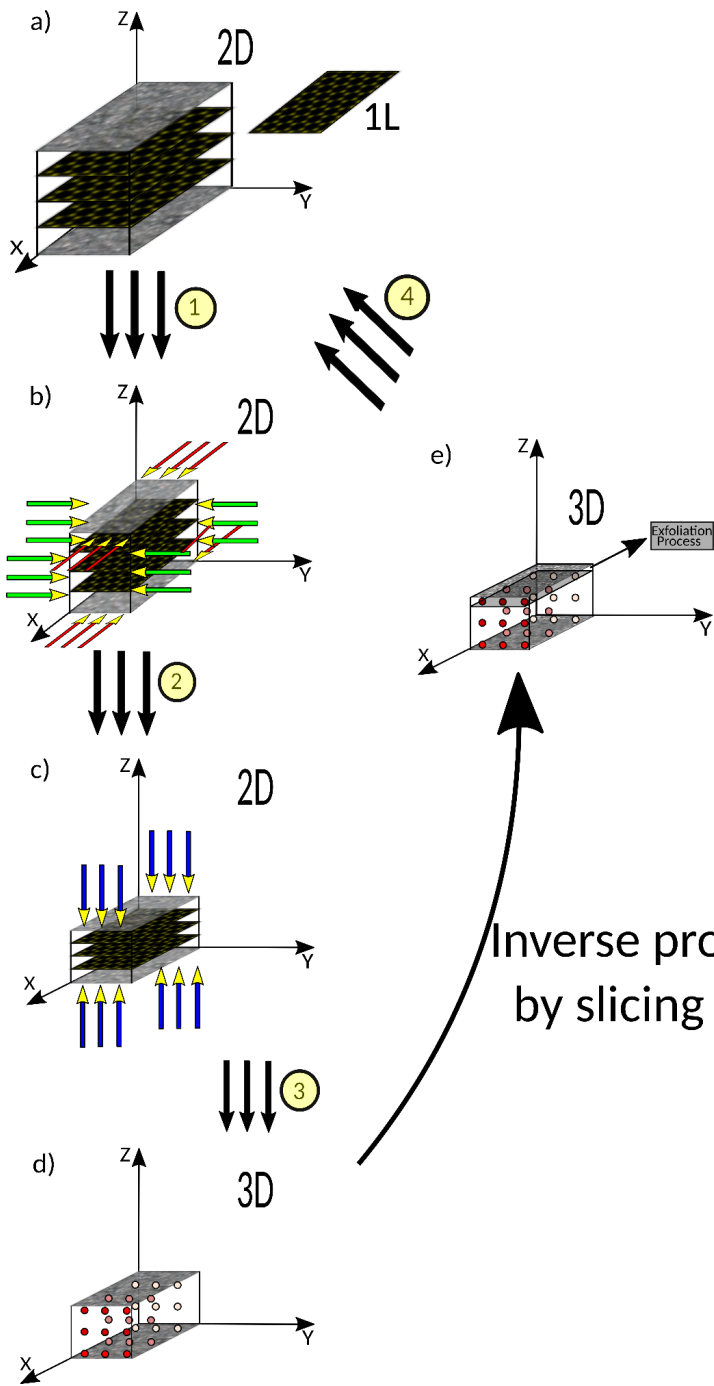

**Figure S1**. Scheme of obtaining 3D structures from 2D ones. (1) 'Sliced' 2D layers from the 3D 'target' structure are compressed in different directions (2) until a stable 3D structure is obtained (3). To test for completeness, the 'inverse' process is carried out, 'slicing' the obtained 3D structure to obtain new 2D ones (4). If the obtained 3D structure is not the desired 'target' the process is repeated using different compressing/slicing parameters/2D structures.

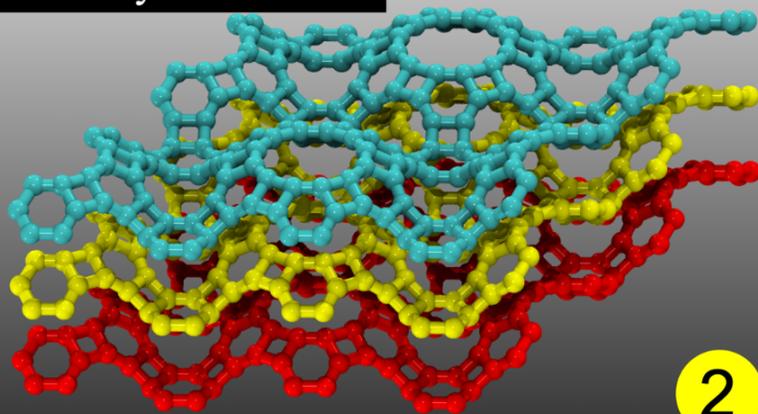

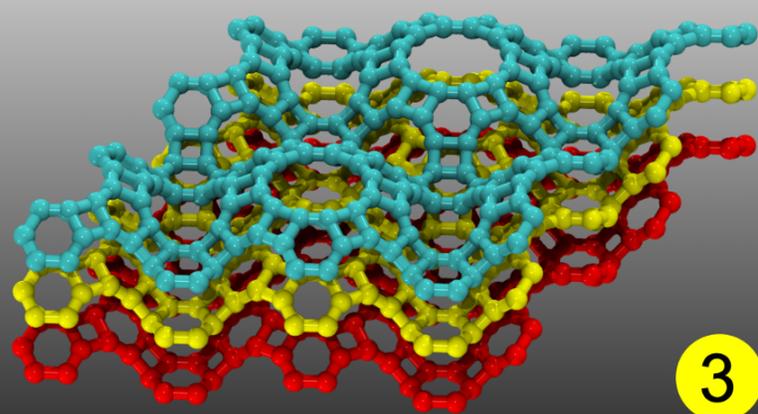

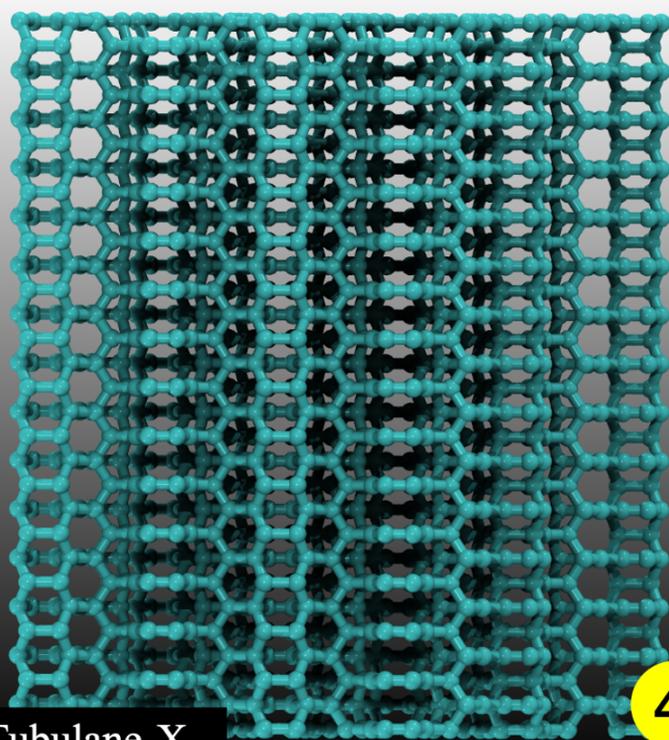

**Figure S2**. Representative snapshots of the structural stages from 2D to 3D ones. Compressed corrugated (buckled) stacked BPC (2). The top (3) and lateral view (4) of the obtained 3D structure, named Tubulane X, which is an unreported new member of the Tubulane family [22].

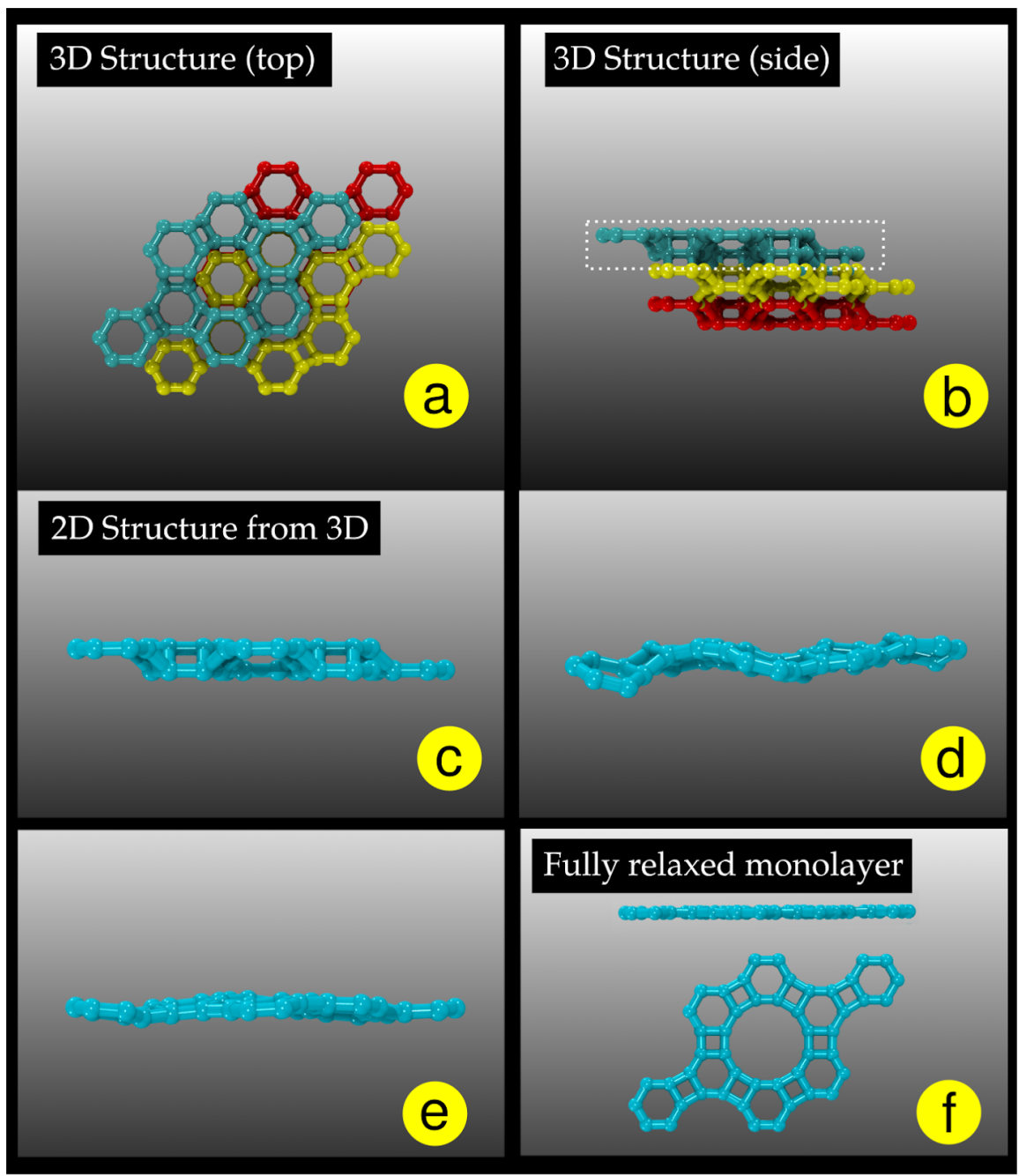

**Figure S3:** 'Reverse' test '. 'Sliced' structures form the obtained 3D bcc-C6 [21] crystal recover the 'parent' 2D BPC.

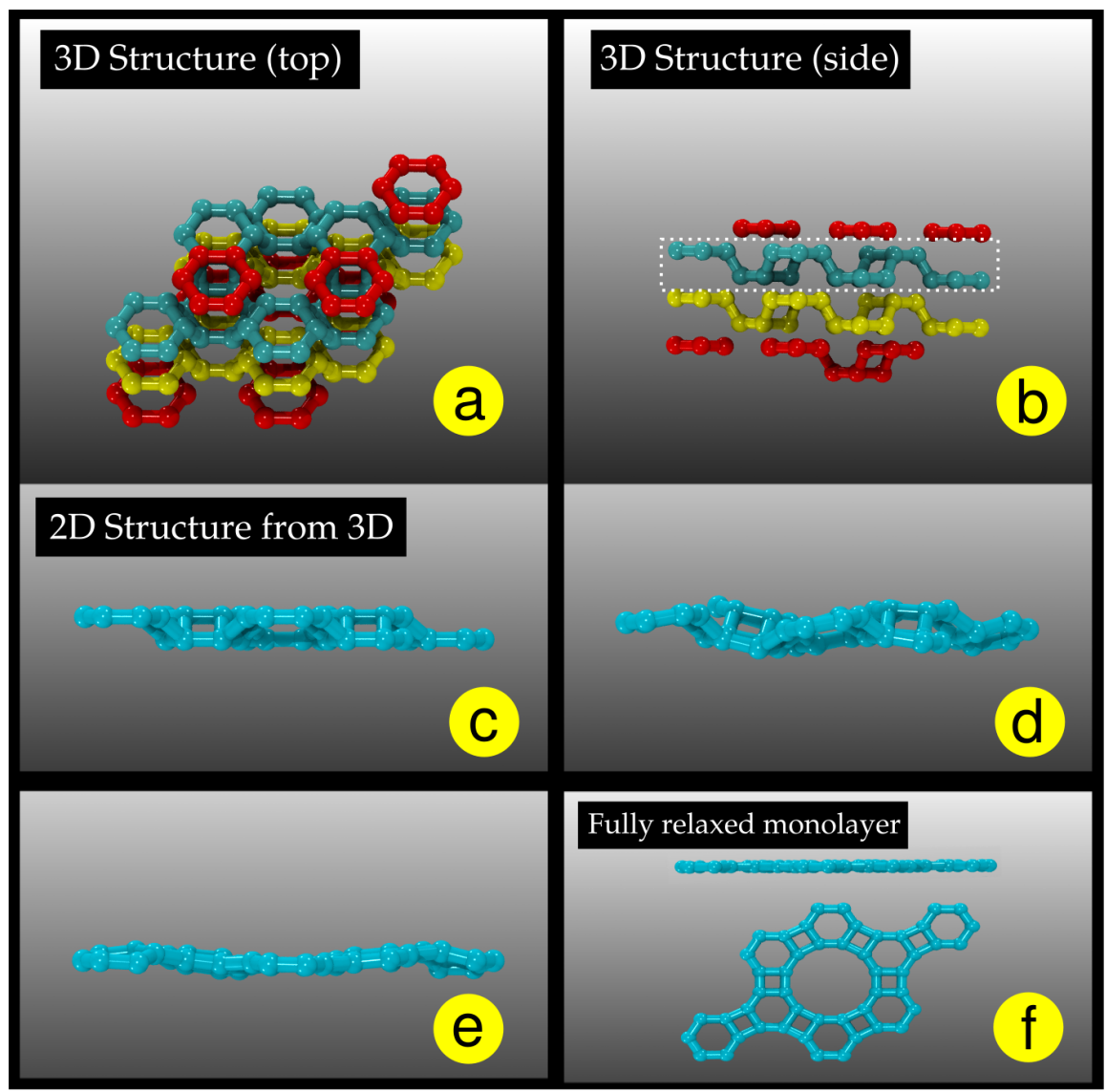

**Figure S4:** 'Reverse' test '. 'Sliced' structures form the obtained 3D Tubulane 12-hexa(3,3) [22] recover the 'parent' 2D BPC.

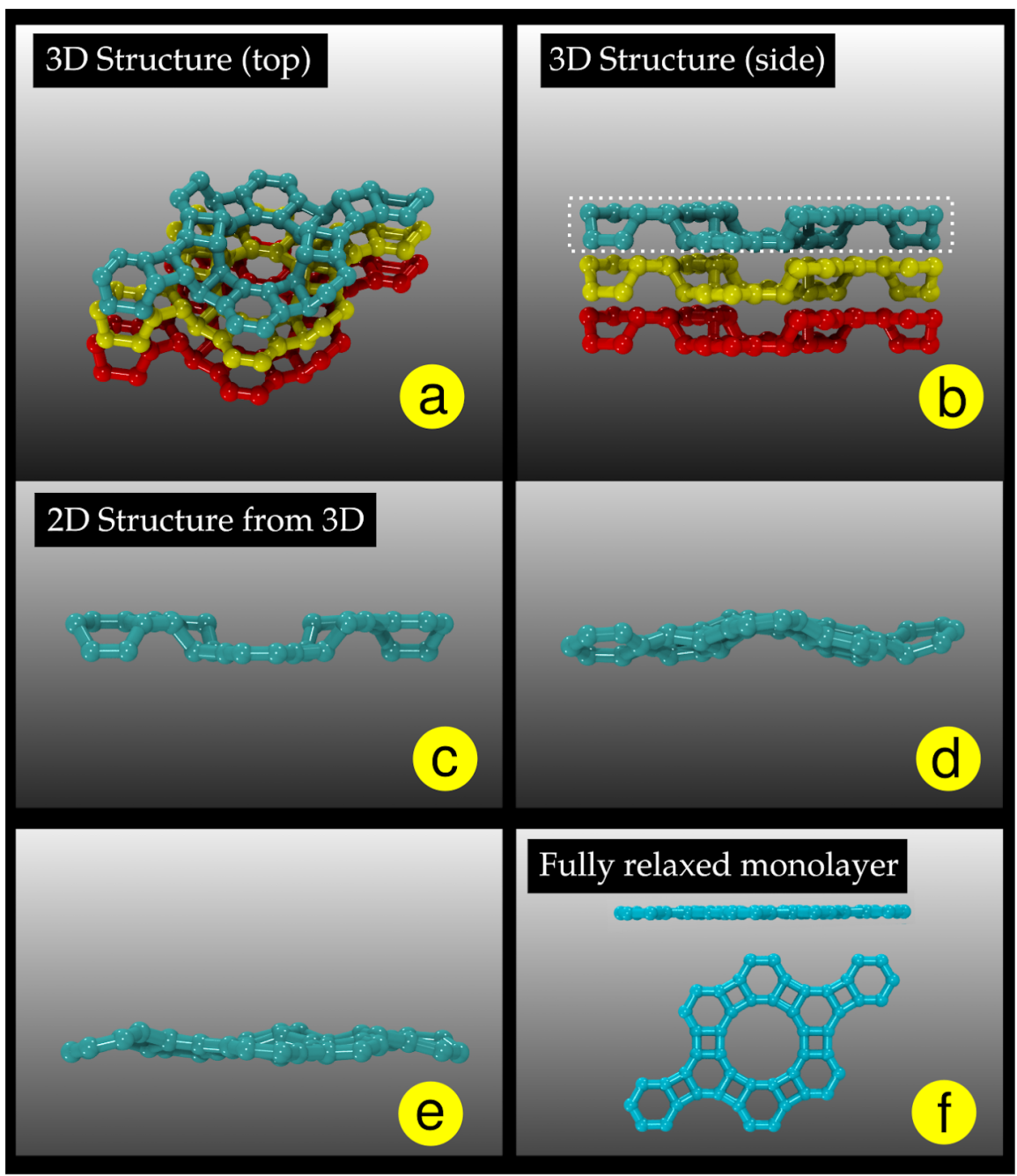

**Figure S4:** 'Reverse' test '. 'Sliced' structures form the obtained Tubulane X crystal recover the 'parent' 2D BPC.

[1] Novoselov KS, Geim AK, Morozov SV, Jiang D, Zhang Y, Dubonos SV, et al. Electric field effect in atomically thin carbon films. Science 2004;306:666–9. https://doi.org/10.1126/science.1102896.

[2] Geim AK, Novoselov KS. The rise of graphene. Nat Mater 2007;6:183–91. https://doi.org/10.1038/nmat1849.

[3] Castro Neto AH, Guinea F, Peres NMR, Novoselov KS, Geim AK. The electronic properties of graphene. Rev Mod Phys 2009;81:109–62. https://doi.org/10.1103/RevModPhys.81.109.

[4] Young RJ, Kinloch IA, Gong L, Novoselov KS. The mechanics of graphene nanocomposites: A review. Composites Science and Technology 2012;72:1459–76. https://doi.org/10.1016/j.compscitech.2012.05.005.

[5] Zhang P, Ma L, Fan F, Zeng Z, Peng C, Loya PE, et al. Fracture toughness of graphene. Nat Commun 2014;5:3782. https://doi.org/10.1038/ncomms4782.

[6] Bizao RA, Machado LD, de Sousa JM, Pugno NM, Galvao DS. Scale effects on the ballistic penetration of graphene sheets. Sci Rep 2018;8:6750. https://doi.org/10.1038/s41598-018-25050-2.

[7] Wang X-Q, Li H-D, Wang J-T. Prediction of a new two-dimensional metallic carbon allotrope. Phys Chem Chem Phys 2013;15:2024–30. https://doi.org/10.1039/c2cp43070c.

[8] Hudspeth MA, Whitman BW, Barone V, Peralta JE. Electronic properties of the biphenylene sheet and its one-dimensional derivatives. ACS Nano 2010;4:4565–70. https://doi.org/10.1021/nn100758h.

[9] Baughman RH, Eckhardt H, Kertesz M. Structure-property predictions for new planar forms of carbon: Layered phases containing $sp^2$ and $sp$ atoms. J Chem Phys 1987;87:6687–99. https://doi.org/10.1063/1.453405.

[10] Fan Q, Yan L, Tripp MW, Krejčí O, Dimosthenous S, Kachel SR, et al. Biphenylene network: A nonbenzenoid carbon allotrope. Science 2021;372:852–6. https://doi.org/10.1126/science.abg4509.

[11] Hu Y, Wu C, Pan Q, Jin Y, Lyu R, Martinez V, et al. Synthesis of γ-graphyne using dynamic covalent chemistry. Nat Synth 2022. https://doi.org/10.1038/s44160-022-00068-7.

[12] Khan K, Tareen AK, Aslam M, Wang R, Zhang Y, Mahmood A, et al. Recent developments in emerging two-dimensional materials and their applications. J Mater Chem C 2020;8:387–440. https://doi.org/10.1039/C9TC04187G.

[13] Novoselov KS, Mishchenko A, Carvalho A, Castro Neto AH. 2D materials and van der Waals heterostructures. Science 2016;353:aac9439. https://doi.org/10.1126/science.aac9439.

[14] Puthirath Balan A, Radhakrishnan S, Woellner CF, Sinha SK, Deng L, Reyes C de L, et al. Exfoliation of a non-van der Waals material from iron ore hematite. Nat Nanotechnol 2018;13:602–9. https://doi.org/10.1038/s41565-018-0134-y.

[15] Balan AP, Puthirath AB, Roy S, Costin G, Oliveira EF, Saadi MASR, et al. Non-van der

Waals quasi-2D materials; recent advances in synthesis, emergent properties and applications. Materials Today 2022. https://doi.org/10.1016/j.mattod.2022.07.007.

[16] Telling RH, Pickard CJ, Payne MC, Field JE. Theoretical strength and cleavage of diamond. Phys Rev Lett 2000;84:5160–3. https://doi.org/10.1103/PhysRevLett.84.5160.

[17] Khaliullin RZ, Eshet H, Kühne TD, Behler J, Parrinello M. Nucleation mechanism for the direct graphite-to-diamond phase transition. Nat Mater 2011;10:693–7. https://doi.org/10.1038/nmat3078.

[18] Xie H, Yin F, Yu T, Wang J-T, Liang C. Mechanism for direct graphite-to-diamond phase transition. Sci Rep 2014;4:5930. https://doi.org/10.1038/srep05930.

[19] Scandolo S, Bernasconi M, Chiarotti GL, Focher P, Tosatti E. Pressure-Induced Transformation Path of Graphite to Diamond. Phys Rev Lett 1995;74:4015–8. https://doi.org/10.1103/PhysRevLett.74.4015.

[20] Bu H, Zhao M, Dong W, Lu S, Wang X. A metallic carbon allotrope with superhardness: a first-principles prediction. J Mater Chem C 2014;2:2751–7. https://doi.org/10.1039/C3TC32083A.

[21] Yin W-J, Chen Y-P, Xie Y-E, Liu L-M, Zhang SB. A low-surface energy carbon allotrope: the case for bcc-C6. Phys Chem Chem Phys 2015;17:14083–7. https://doi.org/10.1039/c5cp00803d.

[22] Baughman RH, Galvão DS. Tubulanes: carbon phases based on cross-linked fullerene tubules. Chem Phys Lett 1993;211:110–8. https://doi.org/10.1016/0009-2614(93)80059-X.

[23] Stewart JJP. Optimization of parameters for semiempirical methods VI: more modifications to the NDDO approximations and re-optimization of parameters. J Mol Model 2013;19:1–32. https://doi.org/10.1007/s00894-012-1667-x.

[24] Gordeev EG, Polynski MV, Ananikov VP. Fast and accurate computational modeling of adsorption on graphene: a dispersion interaction challenge. Phys Chem Chem Phys 2013;15:18815–21. https://doi.org/10.1039/c3cp53189a.

[25] Ribeiro FJ, Tangney P, Louie SG, Cohen ML. Hypothetical hard structures of carbon with cubic symmetry. Phys Rev B 2006;74:172101. https://doi.org/10.1103/PhysRevB.74.172101.

[26] Umemoto K, Wentzcovitch RM, Saito S, Miyake T. Body-centered tetragonal C4: a viable sp3 carbon allotrope. Phys Rev Lett 2010;104:125504. https://doi.org/10.1103/PhysRevLett.104.125504.

[27] Brunetto G, Autreto PAS, Machado LD, Santos BI, dos Santos RPB, Galvão DS. Nonzero Gap Two-Dimensional Carbon Allotrope from Porous Graphene. J Phys Chem C 2012;116:12810–3. https://doi.org/10.1021/jp211300n.